\documentclass[reprint, amsmath,amssymb, aps]{revtex4-2}
\usepackage{graphicx}
\usepackage[export]{adjustbox}% http://ctan.org/pkg/adjustbox
\usepackage{physics}
\usepackage{dcolumn}
\usepackage{bm}
\usepackage{appendix}

\begin{document}

\preprint{APS/123-QED}

\title{Optical transmission of a moving Fabry-Perot interferometer}

\author{Nazar Pyvovar}
\author{Lingze Duan}
\email{ld0003@uah.edu}
\affiliation{Department of Physics and Astronomy, The University of Alabama in Huntsville, Huntsville, AL 35899, USA.}

\date{\today}

\begin{abstract}
Fabry-Perot interferometers have been widely studied and used for well over a century. However, they have always been treated as stationary devices in the past. In this paper, we investigate the optical transmission of a longitudinally moving Fabry-Perot interferometer within the framework of relativity and establish a general relation between the transmission coefficient and the velocity for uniform motions. Several features of the transmission spectrum are analyzed, including velocity-dependent frequency rescaling, the Fabry-Perot transmittance, and the transmission phase near resonance. Special attentions are given to the non-relativistic regime, where application prospects are evaluated. Potential new interferometric schemes, such as velocity-scanning interferometry and hybrid interferometers based on nested configurations, are proposed. Finally, a special case of non-uniform motion is also investigated.
%Fabry-Perot interferometers have been widely studied and used for well over a century. However, they have always been treated as stationary devices in the past. In this paper, we investigate the optical transmission of a longitudinally moving Fabry-Perot interferometer within the framework of relativity and establish a general relation between the transmission coefficient and velocity. In the nonrelativistic regime, the transmittance of the interferometer is found to have a velocity dependence of Airy-function type, which enables the definitions of free spectral range and full-width at half-maximum with respect to velocity. The transmission phase, on the other hand, features a steep linear slope near the resonance condition. These properties of a moving Fabry-Perot interferometer allow it to be used directly for velocity and acceleration measurements. Hybrid interferometric schemes based on nested configurations are proposed as examples of potential applications.
\end{abstract}

\maketitle

\section{\label{sec:level1}Introduction}

The Fabry-Perot interferometer (FPI) is one of the most well-known optical instruments and has been widely used in many fields of science and technology \cite{Hernandez,Vaughan}. In recent years, the advances in hybrid interferometers have created a new modality of utilizing the FPI. These hybrid schemes typically involve nesting the FPI in a "host" interferometer, such as a Michelson interferometer \cite{Abbott,Graf_et_al,Khalil} or a Mach-Zehnder interferometer \cite{Duan}. When the FPI operates under the resonance condition, it folds a long optical path inside and hence extends the effective arm length of the host interferometer, which often leads to much improved phase sensitivities. Such schemes have found tremendous success in gravitational-wave detection \cite{Weiss} and have achieved record-setting strain resolutions in fiber-optic sensing \cite{Duan}. 

\begin{figure}[ht]
\centering
\includegraphics[width=0.95\linewidth]{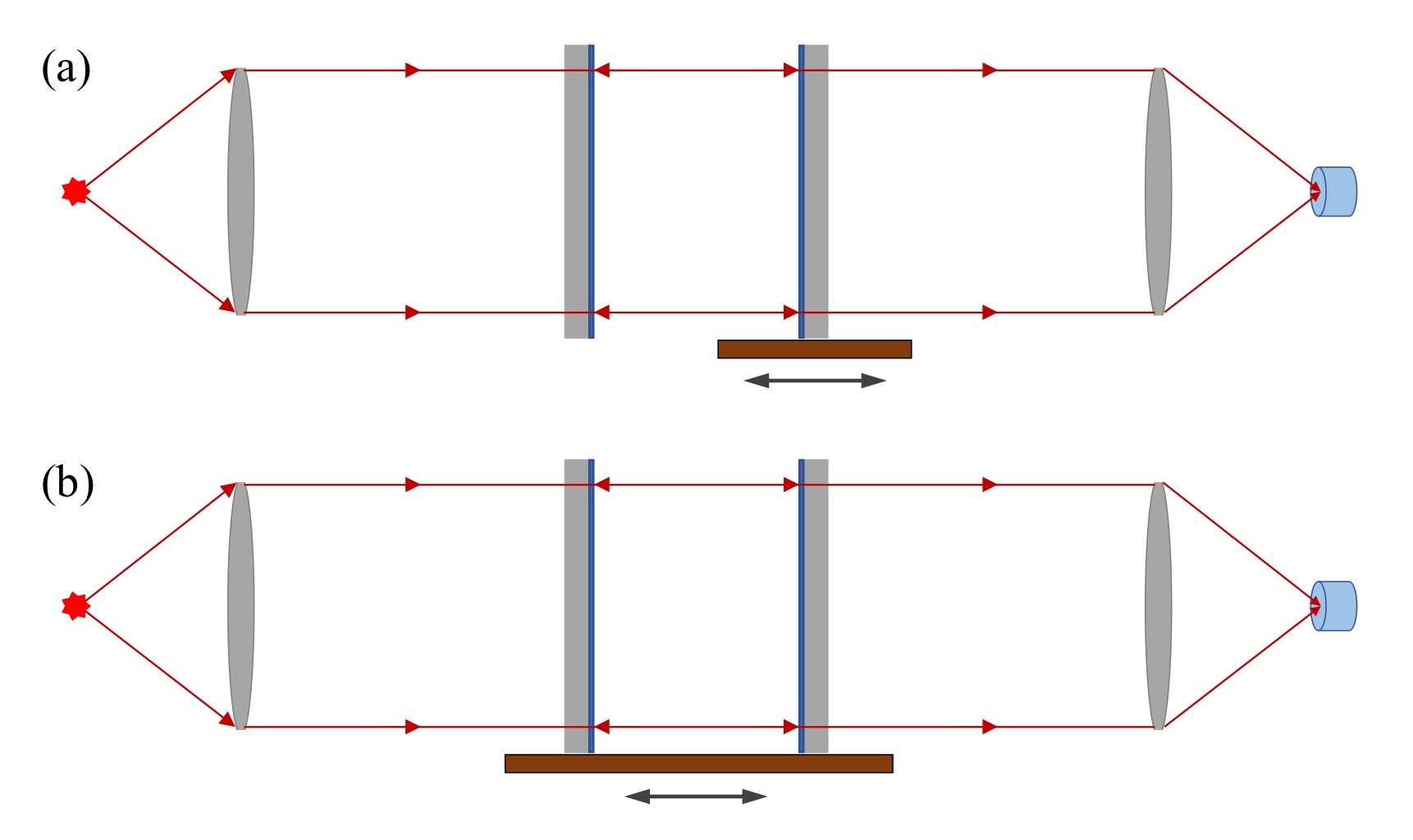}
\caption{The conceptual schemes of (a) a conventional scanning FPI and (b) a moving FPI.}
\label{fig:comparison} 
\end{figure}

Making the FPI a part of another interferometer also provokes an interesting thought: what if the FPI is moving relative to the host interferometer? The question stems from an intuitive rationale: since the FPI is intimately linked to the optical path length in a nested interferometer, any potential phase change induced by a longitudinal motion of the FPI could result in a detectable signal at the output of the host interferometer. Answering this question necessitates a thorough understanding of the transmission properties of an FPI in relative motion with respect to its interrogation system, which usually includes the light source and the detector. Surprisingly, despite being extensively studied for over a hundred years \cite{Fabry_Perot}, the FPI has always been treated as a \textquotedblleft stationary\textquotedblright device. This may sound erroneous at first, because, after all, one of the most common ways of using the FPI is by scanning its mirrors \cite{Ramsay}. However, there is a fundamental distinction between a conventional \textquotedblleft scanning\textquotedblright FPI and a \textquotedblleft moving\textquotedblright FPI to be discussed here, as illustrated in Fig.~\ref{fig:comparison}. In the former case, the transmission properties of the FPI are tuned by changing the positions of the mirrors. Since position is inherently a stationary quantity, the scanning FPI still operates on the premise that the FPI remains stationary with respect to its interrogation system in the lab frame. In the latter case, however, the FPI under concern is rigid and moves as a unit. The relevant question becomes how the states of motion, such as velocity and acceleration, impact the optical properties of the FPI.

This paper presents our study of motion-induced transmission properties for a moving FPI. Before analyzing specific cases, let us first define the general problem under investigation. Consider an FPI formed by two plane mirrors facing each other, with an optical medium set in between. The FPI is situated on a moving stage, which travels along the optical axis as indicated in Fig.~\ref{fig:comparison}(b). A collimated beam of light propagating along the optical axis interrogates the FPI under normal incidence, and the transmitted light is monitored by a photodetector. Both the light source and the detector are located in the lab (rest) frame. 

In the following, we will first examine the optical transmission of the FPI under uniform motion. We will discuss how the velocity dependence of the transmission coefficient impacts the amplitude (or power) and the phase of the transmitted light. Potential schemes of applications are also proposed for velocity and acceleration measurements by means of a moving FPI. Finally, the FPI under arbitrary motions is examined and discussed.

%--------------------------------------------------
\section{Uniform Motion}

%----------------------------
\subsection{General Theory}

We start with the simplest scenario, with the FPI traveling at a constant velocity. In this case, the co-moving frame is an inertial frame, and the transmission coefficient of the FPI can be derived by a simple generalization of the standard multiple-wave superposition approach. 

\begin{figure}[ht]
\centering
\includegraphics[width=0.95\linewidth]{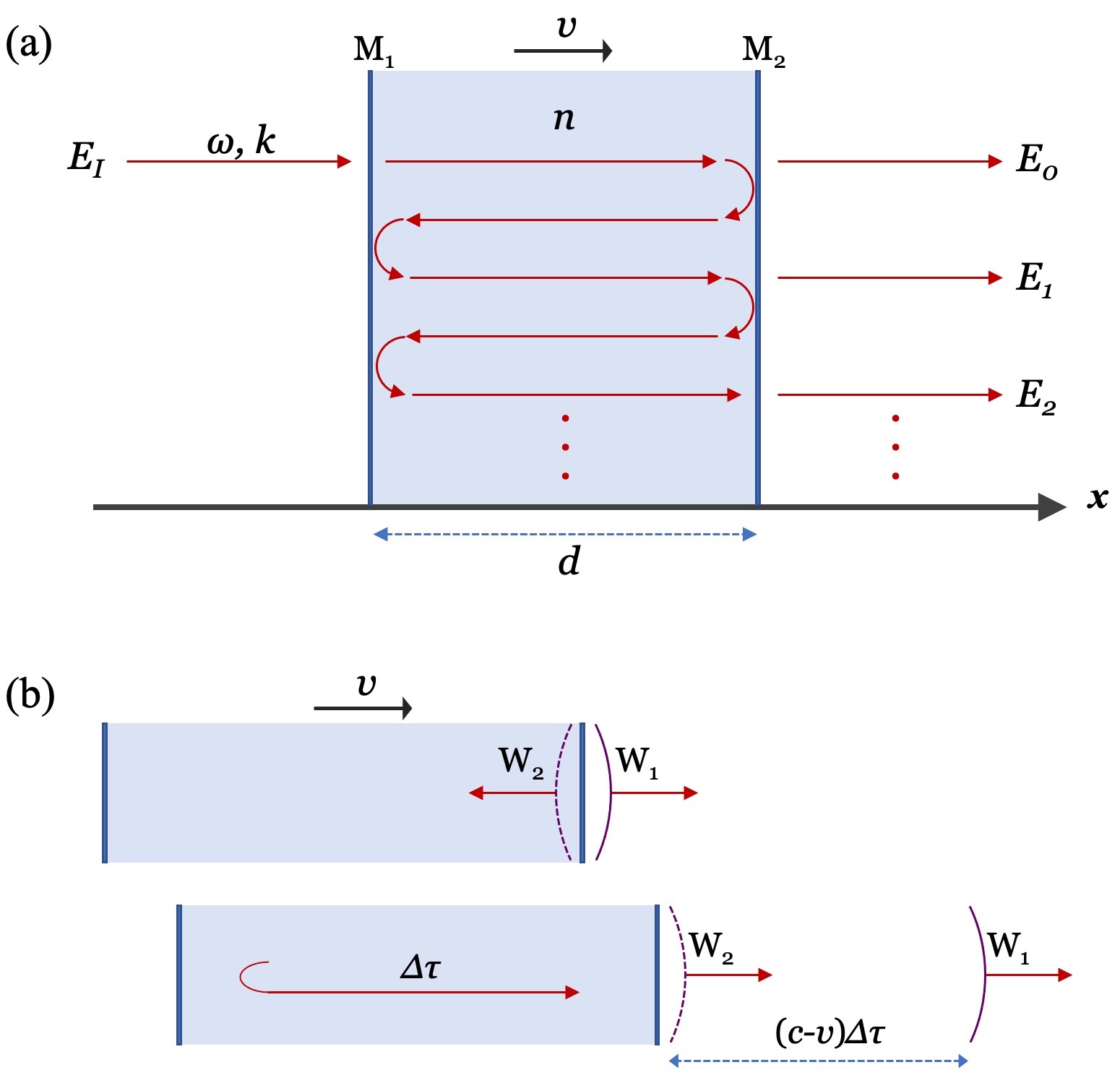}
\caption{(a) The concept of multiple-wave superposition for a uniformly moving FPI. (b) Finding the round-trip phase delay by following the propagation of a wavefront.}
\label{fig:multiwave}
\end{figure}

Consider an FPI of proper length $d$, moving at a constant velocity $v$ along the optical axis $x$, as shown in Fig.~\ref{fig:multiwave}(a). The FPI is made of a homogeneous, isotropic optical medium of refractive index $n$ and is surrounded by vacuum. For simplicity, let us assume that both mirrors of the FPI have the same reflection coefficient $r$ and transmission coefficient $t$. Let us further assume that $r$ and $t$ are independent of velocity. The interrogation wave has a frequency $\omega$ and propagates toward $+x$ along the optical axis. As the wave bounces back and forth between the two mirrors, a series of transmitted field is generated on the exit mirror, as illustrated in Fig.~\ref{fig:multiwave}(a). The overall transmitted field is the superposition of these successive wavefronts. 

Since the incident wave is normal to the mirrors, the electric field $\mathcal{E}$ can be treated as a scalar and in general written as $\mathcal{E}(x,\tau)=E\exp{i(kx-\omega\tau)}$, where $x$ and $\tau$ represent space and time in the lab frame, respectively, $k$ is the wave number in vacuum in the lab frame, and $E$ is the complex amplitude. On the exit mirror of the FPI, the total transmitted field is given by $E_T=E_0+E_1+E_2+...$, where $E_0$, $E_1$, $E_2$, ... are consecutive transmitted wavefronts after each round trip. Assuming a unity incident field $E_I=1$ and neglecting all losses due to absorption, the absolute value of the $N$th wavefront can be expressed as $|E_N|=t^2\cdot r^{2N}$. The phase of $E_N$ is given by $\arg(E_N)=N\Delta \phi$, where $\Delta \phi$ is the round-trip phase shift of the wave. When the FPI is at rest, $\Delta \phi = 2nkd$, and the above superposition leads to the well-known FPI transmission coefficient
\begin{equation}
    T \equiv \frac{E_T}{E_I} = \frac{t^2e^{-inkd}}{1-r^2e^{-2inkd}}.
    \label{T_stationary}
\end{equation}

In the case of a uniformly moving FPI, $\Delta \phi$ is no longer equal to $2nkd$. This is because, for an observer in the lab frame, \textit{i)} the counter-propagating waves inside the FPI no longer share the same frequency and wave number due to the Doppler shifts caused by the motion of the mirrors \cite{Ives,Gjurchinovski}, and \textit{ii)} the counter-propagating waves also travel at different speeds due to the Fizeau's light-dragging effect caused by the moving material \cite{Fizeau, Lorentz, Kuan}. The first effect can be circumvented in the derivation of $\Delta \phi$ by following the propagation of a single wavefront inside the cavity. This is due to the fact that each round trip of the wave involves two bounces on the mirrors, and the induced Doppler shifts exactly cancel out. As the result, all the transmitted wavefronts have the same frequency as the incident wave. Meanwhile, the second condition can be incorporated in the derivation by treating the intracavity material as an anisotropic medium \cite{Chyla,Shen}.  

As illustrated in Fig.~\ref{fig:multiwave}(b), the mirror M\textsubscript{2} splits a wavefront into a transmitted wavefront W\textsubscript{1} and a reflected wavefront W\textsubscript{2}. After W\textsubscript{2} takes a round trip inside the FPI and exits from M\textsubscript{2}, the phase difference between the two wavefronts is the round-trip phase delay $\Delta \phi$. To an observer in the lab frame, the total time the wave spends in a round trip is the sum of its backward and forward propagation times within the FPI, which is given by
\begin{equation}
\Delta \tau = \frac{d'}{(c/n_-) + v} + \frac{d'}{(c/n_+) - v},
\label{Delta_tau1}
\end{equation}
where $c$ is the speed of light in vacuum, $d'$ is the apparent length of the FPI to the observer, $n_\pm$ are the apparent refractive indices of the cavity medium for the forward and the backward propagating waves, respectively, and $v$ is the speed of the FPI. $n_\pm$ can be obtained via the Lorentz transformation of the refractive index \cite{Chyla, Shen},
\begin{equation}
    n_{\pm}=\frac{n \pm \beta}{1 \pm n\beta},
\label{Index_transform}
\end{equation}
where $\beta \equiv v/c$. Note that $d'$ is related to the proper cavity length $d$ through the relation $d' = d/\gamma$, where $\gamma = 1/\sqrt{1-\beta^2}$. It is then straightforward to show that $\Delta \tau$ can be written as 
\begin{equation}
\Delta \tau = \frac{2nd}{c} \gamma.
\label{Delta_tau2}
\end{equation}
During this period of $\Delta \tau$, the FPI also travels forward by a distance $v\Delta \tau$. The total phase difference between the two wavefronts, as illustrated in Fig.~\ref{fig:multiwave}(b), is given by 
\begin{equation}
    \Delta\phi = \omega\Delta\tau - k v\Delta \tau=k(c-v)\Delta\tau.
    \label{Delta_phi1}
\end{equation}
Combining \eqref{Delta_tau2} and \eqref{Delta_phi1} leads to the final expression of the round-trip phase delay
\begin{equation}
    \Delta\phi = 2nkd \sqrt{\frac{1-\beta}{1+\beta}}.
    \label{Delta_phi2}
\end{equation}

This round-trip phase shift is used to replace $\Delta \phi = 2nkd$ in \eqref{T_stationary}, and the transmission coefficient for a uniformly moving FPI can be written as
\begin{equation}
    T = \frac{t^2e^{-i\zeta nkd}}{1-r^2e^{-2i\zeta nkd}},
    \label{T_const_v}
\end{equation}
where $\zeta$ is defined as
\begin{equation}
\zeta = \sqrt{\frac{1-\beta}{1+\beta}}.
\label{zeta}
\end{equation}

Clearly, $\zeta$ represents the impact of the uniform motion on the transmission property of the FPI: rescaling the round-trip phase with a velocity-dependent factor. One may quickly realize that $\zeta$ indicates a relativistic Doppler shift \cite{Ataman}. Its physical meaning becomes clear when the observation is made in the co-moving frame, where the FPI is at rest whereas the light source is moving toward the opposite direction. Since the co-moving frame is an inertial frame and physical properties such as the transmission coefficient is independent of the selection of inertial frames, the transmission of the FPI satisfies
\begin{equation}
    \Tilde{T}=\frac{t^2e^{-in\Tilde kd}}{1-r^2e^{-2in\Tilde kd}},
    \label{T_tilde}
\end{equation}
where the tildes are used to denote quantities in the co-moving frame. A straightforward application of the Lorentz transformation of $k$ yields
\begin{equation}
    \Tilde k=\gamma(k-\frac{v}{c}\frac{\omega}{c})=\zeta k,
    \label{k_transform}
\end{equation}
which leads to the same expression as \eqref{T_const_v}.

%--------------------------------------------------
\subsection{Frequency Rescaling}

The scaling factor $\zeta(v)$ causes some interesting changes to the transmission spectrum of a moving FPI. This becomes clear when we examine the transmittance of the FPI, which takes the form of the well-known Airy function, with an added velocity dependence
\begin{equation}
    |T|^2=\frac{1}{1 + (4\mathcal{F}^2/\pi^2) \sin^2{(\zeta(v)nkd)}},
    \label{Airy_Function}
\end{equation}
where $\mathcal{F} \equiv \pi r/(1 - r^2)$ is the finesse of the FPI. Clearly, this is a generalization of the FPI transmittance for a stationary cavity, which now becomes a special case of \eqref{Airy_Function} with $\zeta(v)=1$ ($v=0$). The resonance condition of the FPI is now given by $\zeta(v)nkd = m\pi$, where $m$ is an integer representing the index of the resonance mode. The condition can be rewritten in terms of wavelength as $m\lambda = 2\zeta(v)nd$ or in terms of optical frequency as $\nu_m = m/[\zeta(v)\Delta \tau_0]$, where $\nu_m$ is the frequency of the $m$th mode and $\Delta \tau_0 = 2nd/c$ is the cavity round-trip time when the FPI is at rest. 

In almost all practical cases, the FPI operates in the nonrelativistic limit, where $\beta\ll 1$. Under this condition, the resonance frequencies are given by
\begin{equation}
    \nu_m = \frac{m}{\Delta \tau_0}(1 + \frac{v}{c}).
    \label{Resonance_Peaks}
\end{equation}
This resonance condition is similar to that of a stationary FPI except for the extra scaling factor $1+v/c$. With a positive velocity, i.e., when the FPI travels in the same direction as the incident light, this scaling factor is greater than 1, which means the mode spacing grows larger than a stationary FPI. The actual frequency shift for the $m$th mode is $mv/\Delta \tau_0 c$. Such a frequency shift can be well within the detectable range even for small velocities, e.g., 1 cm/s. This is because the index $m$ for a typical-sized FPI operating at an optical wavelength is very large ($10^6$\textendash$10^7$).

It is also interesting to point out that the motion of the FPI also causes a rescaling of the transmission linewidth of the FPI by the factor $\zeta(v)$. For example, a positive nonrelativistic velocity would broaden the transmission line by a factor $1+v/c$.

%--------------------------------------------------
\subsection{Velocity-Scanning FPI}

%----------------------------
\subsubsection{Transmittance}

\begin{figure}[ht]
\centering
\includegraphics[width=0.95\linewidth]{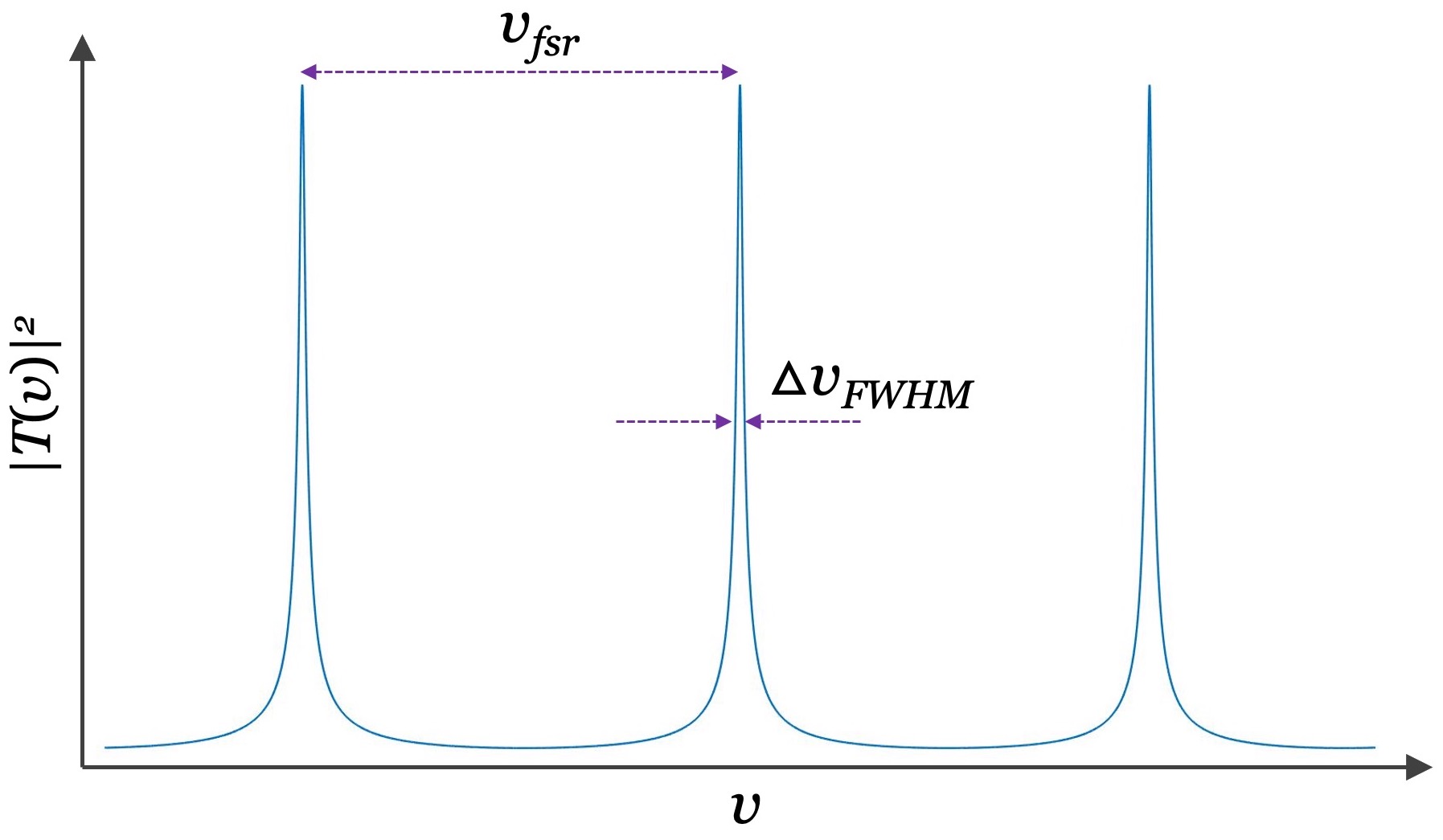}
\caption{The transmittance of a velocity-scanning FPI features periodic transmission peaks versus velocity $v$. The associated FSR and FWHM can be defined accordingly.}
\label{fig:transmittance} 
\end{figure}

Meanwhile, the velocity dependence of FPI transmission leads to a new way to operate the FPI: scanning its longitudinal velocity. In the nonrelativistic regime, the transmittance of a moving FPI can be written as 
\begin{equation}
    |T(v)|^2 = \frac{1}{1 + (2\mathcal{F}/\pi)^2 \sin^2{[nkd(1-v/c)]}}.
    \label{Transmittance}
\end{equation}
$|T(v)|^2$ is a periodic function of $v$, as illustrated in Fig.~\ref{fig:transmittance}. Such a relationship is analogous to that of a scanning FPI, where $|T(d)|^2$ is a periodic function of the cavity length $d$. The resonance peaks appear when the velocity satisfies the condition $v/c = 1-m(\lambda/2nd)$, where $\lambda \equiv 2\pi/k$ is the wavelength. A free spectral range (FSR) $v_{fsr}$ can be defined as the spacing between adjacent resonance peaks, and $v_{fsr}$ is given by
\begin{equation}
    v_{fsr} = \frac{\lambda}{2nd} c = \frac{\lambda}{\Delta \tau_0},
    \label{FSR}
\end{equation}
The linewidth of the resonance peaks is characterized by their full-width at half-maximum (FWHM) $\Delta v_{FWHM}$, which is given by 
\begin{equation}
    \Delta v_{FWHM} = \frac{\lambda}{\Delta \tau_0} \frac{1-r^2}{\pi r} = \frac{v_{fsr}}{\mathcal{F}}.
    \label{FWHM}
\end{equation}

To put the above analysis in a practical context, a one-meter glass ($n=1.5$) FPI interrogated by a laser of $\lambda = 600$ nm would have a FSR $v_{fsr}=60$ m/s. If the FPI has a finesse $\mathcal{F}=1000$, the characteristic width of the resonance peaks is $\Delta v_{FWHM} = 6$ cm/s, which gives the velocity resolution of the FPI as a velocity discriminators. 

%----------------------------
\subsubsection{Transmission Phase}

%Now, let us revisit the question raised at the beginning of the paper: in a hybrid interferometer, where an FPI is used to fold the optical path, what would happen if the FPI begins to move? To answer the question, it is convenient to rephrase it into a more direct question: how does the transmission phase vary near resonance when the FPI has a non-zero velocity? The equivalency between the two questions rests upon the fact that, when an FPI is used to fold the optical path, it always operates on resonance, and when a small deviation is introduced upon the resonance condition, the dominant effect is a phase change \cite{Black}.

The concept of using a moving FPI as a velocity discriminator can also be investigated from the perspective of transmission phase. In general, the transmission coefficient of an FPI can be written as $T = |T|\exp{-i\Phi}$, where $\Phi$ is the transmission phase.  From \eqref{T_const_v}, it is easy to show that $\Phi$ is given by
\begin{equation}
    \Phi = \zeta nkd + \arctan(\frac{r^2 \sin{(2 \zeta nkd)}}{1 - r^2 \cos{(2 \zeta nkd)}}),
    \label{Phi}
\end{equation}
or equivalently,
\begin{equation}
    \tan\Phi=\frac{1+r^2}{1-r^2}\tan (\zeta nkd).
    \label{tan_Phi}
\end{equation}

\begin{figure}[ht]
\centering
\includegraphics[width=0.95\linewidth]{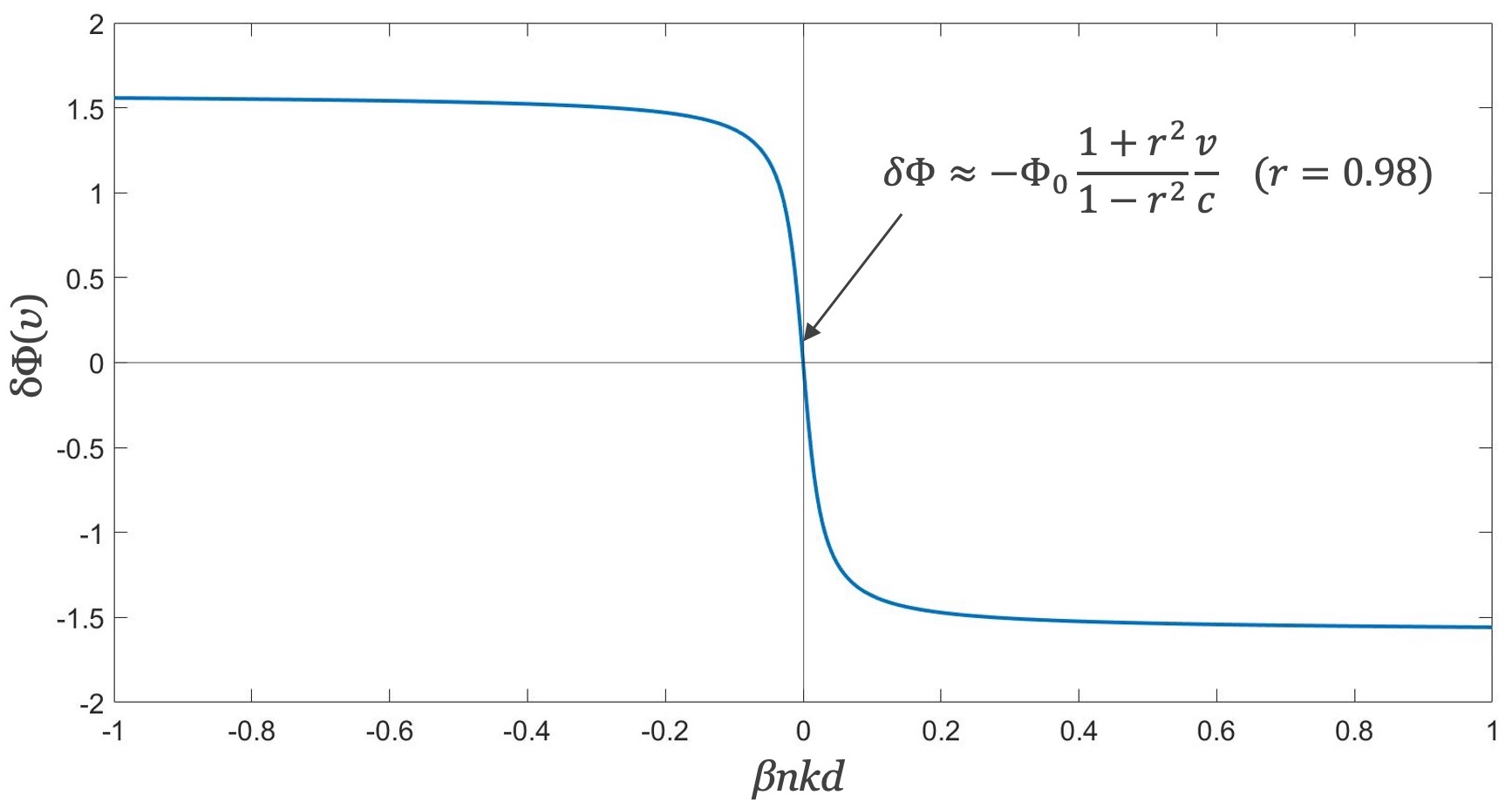}
\caption{The transmission phase of a moving FPI has a near linear velocity dependence when $|\beta nkd|$ is small.}
\label{fig:transmission_phase} 
\end{figure}

Taking $\zeta \approx 1-\beta$ for nonrelativistic motion and using the resonance condition that requires $nkd$ to be an integer multiple of $\pi$, the transmission phase can be rewritten as $\Phi = \Phi_0 + \delta \Phi(v)$, where $\Phi_0 = nkd$ is the steady-state transmission phase when the FPI is stationary. $\delta \Phi(v)$ is a small velocity-dependent phase change caused by the motion and is given by
\begin{equation}
    \delta \Phi(v) = -\beta nkd - \arctan(\frac{r^2 \sin{(2\beta nkd)}}{1 - r^2 \cos{(2\beta nkd)}}).
    \label{delta_Phi}
\end{equation}
In Fig.~\ref{fig:transmission_phase}, $\delta \Phi(v)$ is plotted against $\beta nkd$ with both positive and negative velocities. Apparently, when $|\beta nkd|$ is small, $\delta \Phi(v)$ has a linear dependence over $v$. This linear relation can be found by taking the first-order approximation of \eqref{delta_Phi}, which yields
\begin{equation}
    \delta \Phi(v) \approx -\Phi_0\frac{1+r^2}{1-r^2} \frac{v}{c}.
    \label{delta_Phi_approx}
\end{equation}

We are now ready to revisit the question raised at the beginning of the paper: in a hybrid interferometer, where an FPI is used to fold the optical path, what would happen if the FPI begins to move? When an FPI operating in resonance begins to move longitudinally, albeit \textquotedblleft slowly\textquotedblright, the transmitted light experiences a small phase shift given by \eqref{delta_Phi_approx}. The \textquotedblleft $-$\textquotedblright sign indicates that, when the FPI moves toward the same direction as the optical wave, the phase change is negative. The amount of this phase shift is proportional to the steady-state transmission phase $\Phi_0$, the velocity $v$, as well as the factor $(1+r^2)/(1-r^2)$. Clearly, long cavity length (large $\Phi_0$) and high finesse ($r \rightarrow 1$) leads to high velocity sensitivity of $\delta \Phi(v)$. 

It should be noted that the approximation \eqref{delta_Phi_approx} is only valid when $2\beta nkd \ll 1$, or in other words, $v \ll c/2nkd$. To put this condition in context, for a 1-cm thick glass etalon interrogated at a wavelength of 600 nm, $c/2nkd \approx 1$ km/s, i.e., about three times the speed of sound in the air. Therefore, with cavities of reasonable sizes, \eqref{delta_Phi_approx} can remain valid for most practical cases.

%----------------------------
\subsubsection{Hybrid Interferometers}

\begin{figure}[ht]
    \centering
    \includegraphics[width=0.9\linewidth]{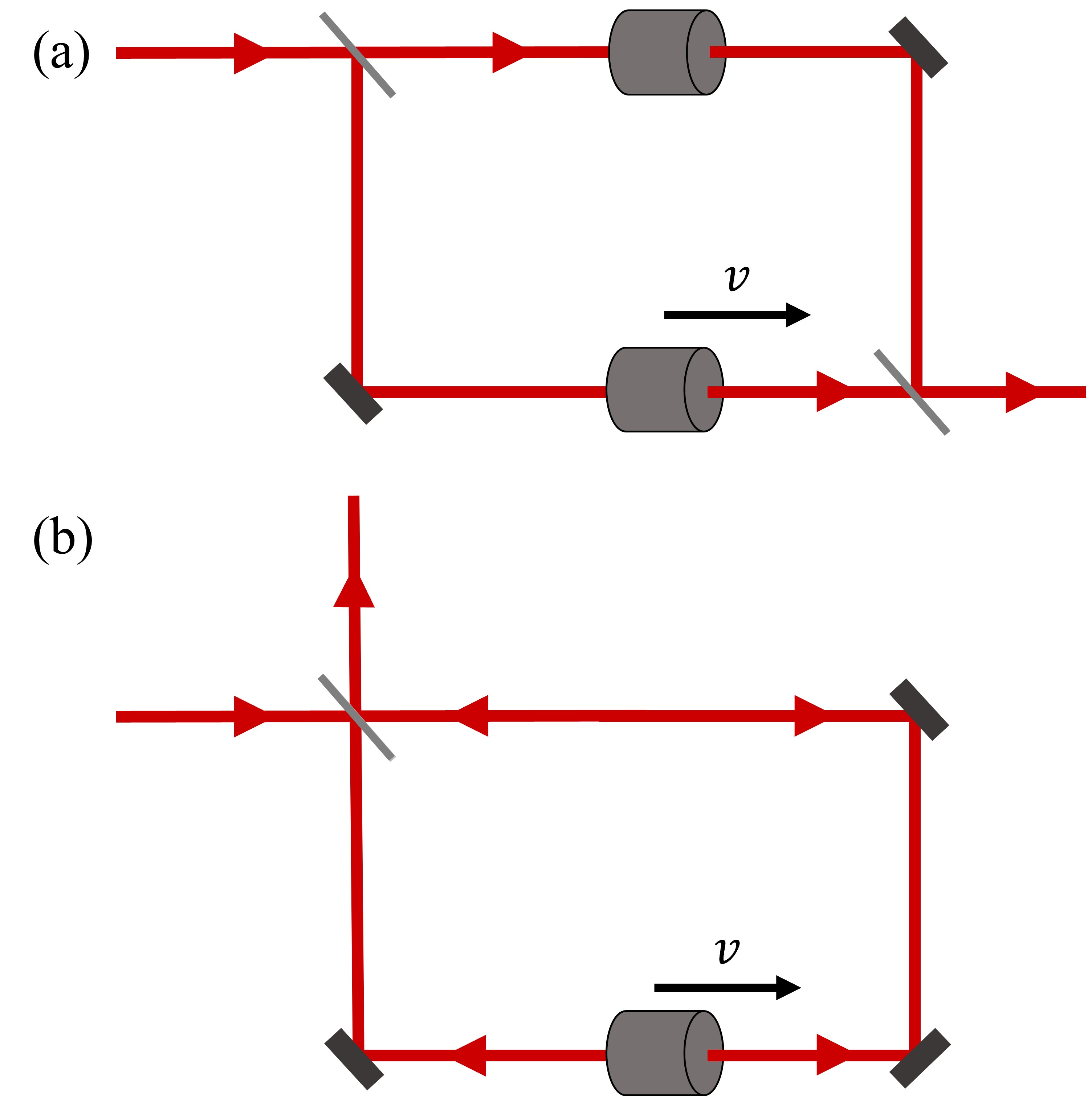}
    \caption{Examples of hybrid interferometers with a moving FPI incorporated: (a) balanced Mach-Zehnder interferometer, and (b) Sagnac interferometer.}
    \label{fig:sagnac}
\end{figure}

The velocity sensitivity of the transmission phase enables highly sensitive velocity measurement using hybrid interferometer configurations. Two possible schemes are shown in Fig.~\ref{fig:sagnac}, and more variations can potentially be conceived. In the first scheme (Fig.~\ref{fig:sagnac}(a)), a balanced Mach-Zehnder interferometer (MZI) is used as the host interferometer, with two identical FPIs serving as optical path multipliers. Such a hybrid configuration has been experimentally demonstrated in fiber-optic systems \cite{Duan}. Now, if one of the FPIs begins to move longitudinally while the other remains stationary, an extra phase difference is generated between the two arms according to \eqref{delta_Phi_approx}, causing a detectable signal at the output of the MZI.

In the second scheme, a single FPI is integrated into a Sagnac interferometer (SI), as shown in Fig.~\ref{fig:sagnac}(b). Under the steady state with the FPI at rest, the clockwise and the counterclockwise circulating beams experience the same optical path inside the FPI. When the FPI moves toward one direction, however, this symmetry between the two directions is broken, and a phase shift $\delta \Phi_{SI}$ is created between the two beams at the output port of the SI. Note that the velocity sensitivity of the SI configuration is twice as large as the MZI configuration:
\begin{equation}
    \delta \Phi_{SI} = 2\Phi_0\frac{1+r^2}{1-r^2} \frac{v}{c}.
    \label{Sagnac}
\end{equation}
Therefore, for velocity and acceleration measurements using moving FPIs, the simpler SI configuration may be a more preferable scheme.

%--------------------------------------------------
\section{Adiabatic Non-Uniform Motion}

For non-uniformly moving FPIs, a general analysis is too involved to fit into the scope of this work. Here, we only focus on a relatively simple case: adiabatic non-uniform motion (ANUM). The term \textquotedblleft adiabatic\textquotedblright refers to the condition where the change of velocity is so \textquotedblleft slow\textquotedblright that, at any moment of time, the FPI can be approximately treated as being in a state of uniform motion. Physically, this requires the change of velocity during the effective \textquotedblleft storage time\textquotedblright of the FPI to be less than what the FPI can discriminate. Analytically, we can describe this condition simply as $(dv/dt) \tau_{cav} < \Delta v_{FWHM}$, where $\tau_{cav}$ is the cavity storage time for an FPI and is typically given by $\tau_{cav} = 2nd \mathcal{F}/(\pi c)$ \cite{Lawrence}. Using \eqref{FWHM}, the ANUM condition can be written as
\begin{equation}
a < \frac{\pi \lambda}{\Delta \tau_0^2 \mathcal{F}^2},
\label{ANUM}
\end{equation}
where $a \equiv dv/dt$ is acceleration. Again, let us put this relation into context by applying the following typical values: $d = 0.1$ m, $n = 1$, $\mathcal{F} = 1000$, and $\lambda = 600$ nm. The corresponding ANUM condition is $a < 4 \times 10^6$ m/s\textsuperscript{2}, which is about $4 \times 10^5$ times the earth gravity acceleration. Therefore, for most foreseeable applications, the ANUM condition should be well satisfied.

Under the ANUM condition, the transmission coefficient of the FPI adiabatically follows the change of the velocity such that
\begin{equation}
    T(\tau) = \frac{t^2e^{-i\zeta(\tau) nkd}}{1-r^2e^{-2i\zeta(\tau) nkd}},
    \label{T_adiabatic}
\end{equation}
where $\zeta(\tau)$ is defined as
\begin{equation}
\zeta(\tau) = \sqrt{\frac{1-v(\tau)/c}{1+v(\tau)/c}}.
\end{equation}

%--------------------------------------------------
\section{Conclusion}
In conclusion, we have analyzed the transmission coefficient of a moving FPI and have shown that it is closely related to the velocity of the FPI. In the nonrelativistic regime where $\beta \ll 1$, the transmittance of the FPI has a velocity dependence that resembles the cavity-length dependence of a scanning FPI or the wavelength dependence of a Fabry-Perot filter. Corresponding concepts of FSR and FWHM can be defined with respect to velocity. Meanwhile, the transmission phase of the FPI is shown to feature a sharp linear slope around the resonance velocities. These properties of the transmission coefficient allow a moving FPI to be used directly for velocity and acceleration measurements. As examples, hybrid interferometric schemes based on nested MZI-FP and SI-FP configurations are proposed. We hope that this work can offer a new perspective to the century-old device FPI.

\section{Acknowledgements}
This work has been supported in part by the National Science Foundation under Grant ECCS-1606836.

\nocite{*}
\bibliography{apsbib}

\end{document}